# Sunflower nested core collections for association studies and phenomics

Marie Coque [1], Sébastien Mesnildrey[2], Michel Romestant [3], Bruno Grezes-Besset [4], Félicity Vear [5], Nicolas B. Langlade [6], and Patrick Vincourt [6]

[1] SYNGENTA Seeds S.A.S.
[2] EURALIS Semences
[3] RAGT Semences
[4] BIOGEMMA S.A.S.
[5] INRA, UMR INRA-Université Blaise Pascal 1095 234, Avenue du Brezet F-63000 Clermont-Ferrand
[6] INRA, Laboratoire des Interactions Plantes Micro-organismes (LIPM), UMR CNRS-INRA 2594-441, F-31320 Castanet Tolosan, France

## ABSTRACT

In order to develop association studies and to improve the phenotypic description for abiotic and biotic stress related traits, nested core collections of 48, 96, 144 and 384 sunflower lines were built from a set of 752 diverse, public or private accessions. These 752 lines were been genotyped with 51 SSR markers covering the genetic map (3 markers/linkage group). We then used MSTRAT software (Gouesnard et al., 2000) to construct 4 nested core collections as follows: we built a first core with 48 public lines and a kernel of 12 selected entries, accounting for 47% of the total diversity. This short core collection was then used as a kernel to define a second core with 96 public entries, accounting for 59% of the total diversity. Finally the private entries were added to build a core collection of 144 and 384 entries, accounting for 78% and 100% of the total diversity, respectively. The INRA lines belonging to the "96" core collection are available for research to any public institution under a Material Transfer Agreement.

## INTRODUCTION

The development of association studies is becoming very popular to discover relationships between genetic polymorphism and functional variability, either on candidate genes or genome-wide, in human, animal and plant genetics (e.g. Sanguineti et al., 2007, Flint-Garcia et al., 2005). Although less robust than establishing such a relationship through Linkage Mapping in a recombinant population, this approach makes it possible to scan a wider diversity.

As a lot of genomic resources have been or will be developed for crop species, including sunflower, the pertinence and the accuracy of phenotypic description is becoming a potential strong limiting factor when association studies are developed. Thus it appeared useful to establish core collections or panels of different sizes, to establish on a small or medium size panel a correlation between a low throughput phenotyping method and a higher throughput, more adapted to phenotyping of large core collections.

This paper describes the method we used to build nested core collections of different sizes from 752 sunflower inbred lines, and to provide the list of two nested collections (48 and 96 lines) containing public lines or INRA lines made available for the scientific community.

## MATERIALS AND METHODS

**Genetic material:**

The initial genetic material used to build the core collections comprises:
- 350 inbred lines collected or created by INRA at Clermont-Ferrand (F.VEAR, P.LECLERCQ) and at Montpellier (H.SERIEYS, G.PIQUEMAL), and selected by F.VEAR according to their *a priori* phenotypic diversity and to their interest for breeding, from a total list of 2300 lines maintained by INRA.

- 500 private, inbred lines listed by three seed companies companies (RAGT Génétique, SOLTIS, SYNGENTA Seeds), some being elite lines, used as parental lines in hybrid development, the others being created as introgression of wild *Helianthus* ecotypes into an elite line.

**Molecular data:**

All the 850 entries were genotyped with 51 single locus and highly robust SSR markers by the three seeds companies involved in the project. A detailed list of these SSR is available at http://lipm-helianthus.toulouse.inra.fr/Web/core/marker_core.xml .

**Computations:**

We firstly chose a set of 12 inbred lines for their *a priori* diversity and their interest as resources for different research purposes (table 1). We checked whether this choice would affect the diversity of the core collections in locating these lines against the first axis of a Correspondence Analysis made on a table of qualitative data (presence/absence) having the number of genotypes as first dimension, and {locus*alleles} as second dimension.

Then we used MSTRAT (Gouesnard et al., 2000) on qualitative, molecular data, to define the following nested core collections:
- 48 lines with the constraint of including the kernel constituted by the 12 lines listed in Table 1 and of including only public or INRA lines,
- 96 lines with the constraint of including the core collection of 48 lines, and of including only public or INRA lines,
- 144 lines with the constraint of including the core collection of 96 lines, and no other constraint,
- 384 lines with the constraint of including the core collection of 144 lines, and no other constraint.

For each step, 30 iterations and 30 replicates were used, and as recommended by the authors of MSTRAT, we chose the most represented accessions within the 30 replicates.

| Line | Type | Code | Origine | from |
|---|---|---|---|---|
| SF056 | B | FU | Selection from Romanian origins | INRA |
| SF085 | B | CD | HA89 | USDA |
| SF107 | B | 92A6 | *H. argophyllus* introgression | INRA |
| SF109 | B | 2603 | Moroccan population | INRA |
| SF193 | B | XRQ | HA89 * Progress (Russian open pollinated variety) | INRA |
| SF268 | R | RHA266 | RHA266 | USDA |
| SF302 | R | PAC2 | PET1 restorer from H.petiolaris, HA61 | INRA |
| SF306 | R | PAZ2 | PET1 restorer from H.petiolaris, AD66, HA61, Zambia population | INRA |
| SF310 | R | PST5 | Recurrent selection for *Sclerotinia* head-rot resistance | INRA |
| SF317 | R | 83HR4 | Russian origin * RHA274 | INRA |
| SF326 | R | PSC8 | Recurrent selection for *Sclerotinia* head-rot resistance | INRA |
| SF332 | R | RHA274 | RHA274 | USDA |

Table 1: Lines included as the kernel in the 48 core collection built with MSTRAT. 'Line' indicates the line code used for the genotyping, 'Type' indicates if lines restore the male sterility on PET1 cytoplasm (R) or maintain the PET1 cytoplasmic sterility (B), Orig1 indicates the line pedigree.

Finally we checked with STRUCTURE (Pritchard et al., 2000) whether the groups identified by this MSTRAT were correctly represented.

**RESULTS**

**From molecular data:**

As the lines were described as homozygous, we expected to find only a few heterozygous loci from the SSR data. Still, about 100 accessions, presented more than 10% of heterozygous loci according to SSR data. We discarded these lines for further analysis to get the highest confidence in genotypes – as described by SRR – of the resulting core collections.

**Effect of the kernel content on the richness of core collections:**

The 12 lines chosen as members of the kernel in the first step of MSTRAT represented high diversity. They appeared as quite divergent for the first two axes of the Correspondence Analysis (figure 1). As previously found (Tang and Knapp, 2003), the two historic US lines HA89 and RHA274 appeared to be the most divergent within this set of oilseed lines. The group of "R" lines – i.e. restoring the male sterility on PET1 cytoplasm - seemed to be divided into two classes (R1, R2). Some "B" lines – i.e. maintaining the PET1 cytoplasmic sterility- , such as 2603 or 92B6 which includes some *Helianthus argophyllus* background, seemed to be related to the "R2" group. The lines XRQ and PSC8, FU and PAZ2, RHA266 and PAC2 are parental lines of RIL populations developed by F.VEAR at INRA Clermont-Ferrand: together, these RIL populations should account for a large part of the genetic variability, as they reflect the polymorphism exhibited when crossing "B" with "R2", "B" with "R1, "R1" with "R2 respectively.

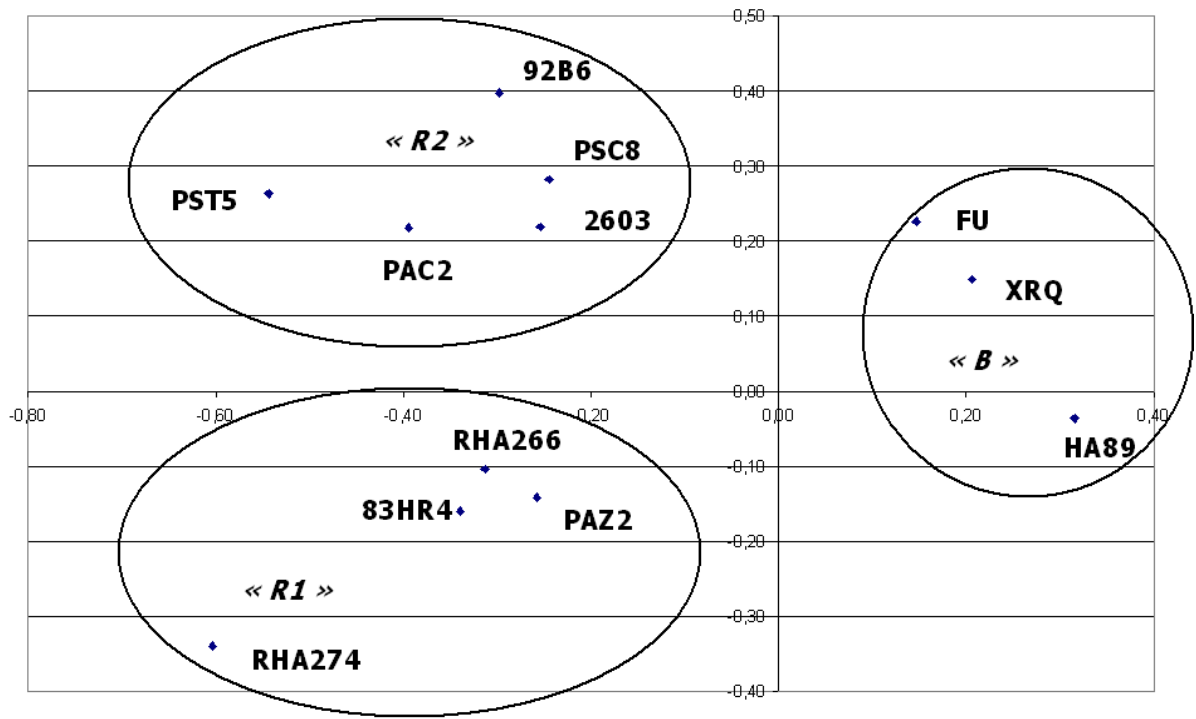

Figure 1: Representation of the lines included in the "kernel" used to build the core collections with MSTRAT, in the two first axis of the Correspondence Analysis.

Another way to evaluate the effect of the kernel content on the richness of the core collection is to compare the score of the kernel with the score of the best core collection of 12 lines within either the set of public and INRA lines, or the total set of lines. As shown in table 2, selecting the kernel dropped the richness only by 25-30% compared to the best "12" core collections.

|  | Kernel | Best core |
|---|---|---|
| Public and INRA | 9,0% | 11,9% |
| All | 6,1% | 8,5% |

Table 2: Comparison of the relative richness associated with the initial kernel of 12 lines with the score of the best 12 subset in the set of public or INRA lines, and in the global set.

**Core collections:**

The first core with 48 public lines including the previously described kernel of 12 entries, accounts for 47% of the total diversity. The core collection of 96 entries, which used the "48" core collection as kernel, accounts for 59% of the total diversity.
The lists of "48" and "96" core collections are provided at http://lipm-helianthus.toulouse.inra.fr/Web/core/Core_collections_list.html
Allowing lines originated from the private seed companies to be added, led to core collections of 144 and 384 entries that account for 78% and 100% of the total diversity, respectively. Indeed, a 100% richness of diversity was obtained with only 220 lines. The lines made by introgression of wild background into sunflower line are greatly explaining the contribution of the private lines.

When we look at the repartition of the members of the "48" or "96" core collection between the 10 classes defined by STRUCTURE, it appears that only 8 out of the 10 classes are represented. For the "144" core collection, which included more diverse genetic backgrounds, all the 10 classes are represented.

## DISCUSSION

As genomic resources are now available for non-model crop plants, the phenotyping activity has become a strong challenge for the workflow efficiency, when linking sequence polymorphisms to the functional evidences. For traits such as yield or adaptation to the abiotic environment, there is no a simple criterion to measure, with a high throughput approach, in the field and/or on a large set of genotypes and environment. Building nested core collections of different sizes will make it possible 1) to establish correlations, on a short set on genotypes chosen for their ability to represent a significant part of the polymorphism, between low throughput physiological criteria and higher throughput indirect measurements, 2) to apply these higher throughput indirect measurements for association studies implying a wider genetic diversity.

These core collections need certainly to be improved for the important gap between the 96 and 144 collections and for the type of data used to build them.
The gap in richness between the "96" core collection and the "144" core collection is mainly due to the introgression lines provided by the seed companies. Introgression lines created by H.Serieys at INRA Montpellier from different wild species is going to be analyzed with the aim to enrich the "96" core collection.
In addition to the marker data, qualitative or quantitative information can be used to build the core collections with the MSTRAT algorithm. The qualification of core collections can be improved in an iterative process where phenotypes recorded on primary core collections are progressively used to check whether there is enough diversity for a particular trait of interest and thus develop *ad hoc* core collections.

**Acknowledgments:** We thank Brigitte GOUESNARD (INRA, UMR DIA-PC, Montpellier) for her helpful comments about MSTRAT.